---------------------------------------------------
\documentstyle [12pt] {article}
\begin{document}
\def\tr{{\rm tr}}
\font\el=cmbx10 scaled \magstep2
%\font\small=cmr8

\def\npar{\indent\hskip 20pt}

%\doublespace

{\obeylines
\hfill CLNS 92/1158
\hfill IP-ASTP-13-92}

\vskip 0.5 cm

\centerline {{\el Chiral Lagrangians for Radiative Decays}}
\centerline {{\el of Heavy Hadrons}}

\bigskip
\bigskip

\centerline{\bf Hai-Yang Cheng$^a$, Chi-Yee Cheung$^a$, Guey-Lin Lin$^a$,}
\centerline{\bf Y. C. Lin$^b$, Tung-Mow Yan$^{a,c}$, and Hoi-Lai Yu$^a$}

\medskip
\centerline{$^a$ Institute of Physics, Academia Sinica, Taipei,}
\centerline{Taiwan 11529, Republic of China}

\medskip
\centerline{$^b$ Physics Department, National Central University,
Chung-li,}
\centerline{Taiwan 32054, Republic of China}

\medskip
\centerline{$^c$ Floyd R. Newman Laboratory of Nuclear Studies, Cornell
University}
\centerline{Ithaca, New York 14853, USA {\it (permanent address)}}

\bigskip
\bigskip

\centerline{\bf Abstract}
{\small

The radiative decays of heavy mesons and heavy baryons are studied in a
formalism which incorporates both the heavy quark symmetry and the chiral
symmetry.  The chiral Lagrangians for the electromagnetic interactions of
heavy hadrons consist of two pieces: one from gauging electromagnetically
the strong-interaction chiral Lagrangian, and the other from the anomalous
magnetic moment interactions of the heavy baryons and mesons.  Due to the
heavy quark spin symmetry, the latter contains only one independent
coupling constant in the meson sector and two in the baryon sector.
These coupling constants only depend on the light quarks and can be
calculated in the nonrelativistic quark model.  However, the charm quark is
not heavy enough and the contribution from its magnetic moment must be
included.  Applications to the radiative decays $D^\ast \rightarrow D
\gamma~,~B^\ast \rightarrow B \gamma~,~ \Xi^\prime_c \rightarrow \Xi_c
\gamma~, \Sigma_c \rightarrow \Lambda_c \gamma$ and $\Sigma_c \rightarrow
\Lambda_c \pi \gamma$ are given.  Together with our previous results on the
strong decay rates of $D^\ast \rightarrow D \pi$ and $\Sigma_c \rightarrow
\Lambda_c \pi$, predictions are obtained for the total widths and branching
ratios of $D^\ast$ and $\Sigma_c$.  The decays $\Sigma^+_c \rightarrow
\Lambda^+_c \pi^0 \gamma $ and $\Sigma^0_c \rightarrow \Lambda^+_c \pi^-
\gamma $ are discussed to illustrate the important roles played by both the
heavy quark symmetry and the chiral symmetry.

 }

\vfill\eject

\noindent{\bf I.~~Introduction}
\bigskip

Mass differences are generally small among the different spin multiplets of
the ground state heavy mesons and heavy baryons which contain a heavy
quark.  This is a consequence of the heavy quark symmetry [1,2] of QCD.  As
a result of the small available phase space, the dominant decay modes for
many of these heavy particles are strong decays with one soft pion emission
and/or radiative decays.  Prominent examples are $D^\ast , B^\ast$ and
$\Sigma_c$ among the heavy particles already observed.  As none of the
absolute widths for these decays has been measured experimentally, it is
important to have a single framework for treating the strong and radiative
decays of these particles.  It will be then possible to test the
predictions on branching ratios of various decay modes with available data.
An ideal theoretical framework for studying these decays is provided by the
formalism recently developed to combine the heavy quark symmetry and the
chiral symmetry of light quarks [3,4,5,6,7,8].  When supplemented by the
nonrelativistic quark model, the formalism determines completely the low
energy dynamics of heavy hadrons.  Among other things, the strong decays
are treated in detail in Ref.[3].  The radiative decays are the subject of
the present work.

The formalism of Ref.[3] is easily extended to incorporate the
electromagnetic field.  The electromagnetic interactions of heavy hadrons
consist of two distinct contributions: one from gauging electromagnetically
the chirally invariant strong interaction Lagrangians for heavy mesons and
baryons given in Ref.[3], and the other from the anomalous magnetic moment
couplings of the heavy particles.  The heavy quark symmetry reduces the
number of free parameters needed to describe the magnetic couplings to the
photon.  For the ground state mesons, there is only one undetermined
parameter, and there are two for the ground state heavy baryons. All these
three parameters are related simply to the magnetic moments of the light
quarks in the nonrelativistic quark model.  However, the charmed quark is
not particularly heavy ($m_c \simeq 1.6$ GeV), and it carries a charge of
$\frac{2}{3} e$.  Consequently, the contribution from its magnetic moment
cannot be neglected.

In the nonrelativistic quark model, all the magnetic moments of hadrons are
due to those of the constituent quarks.  Thus, two of the $1/m_Q$
corrections can easily be taken into account.  The first is to remove the
magnetic moment terms of the heavy hadrons arising from the minimal
couplings to the electromagnetic field.  The second is to include the
contributions from the magnetic moment of the heavy quark.

In Sections II and III we present for heavy mesons and heavy baryons,
respectively, the details of the formalism and related considerations
including the SU(3) flavor symmetry breaking due to light quark mass
differences.

In Section IV we consider applications to the radiative decays of charmed
mesons and charmed baryons.  Some examples are $D^\ast \rightarrow D \gamma
 ~,~ \Xi^\prime_c \rightarrow \Xi_c \gamma ~,~ \Sigma_c \rightarrow
\Lambda_c \gamma $ and $\Sigma_c \rightarrow \Lambda_c \pi \gamma $.  Among
these, perhaps the results for the $D^\ast \rightarrow D \gamma$ decays are
the most interesting.  Experimentally, the most recent CLEO II data [9] on
the branching ratios for $D^{\ast +}$ and $D^{\ast 0}$ differ significantly
from those listed in PDG (1992) [10].  Theoretically, when combined with
our predictions for the strong decays $D^\ast \rightarrow D \pi $ given in
Ref.[3], we are able to obtain the branching ratios for the $D^\ast$ decays
in the same theoretical framework.  Agreement is excellent between theory
and experiment.  This is very encouraging.  Although our predicted total
width for $D^{\ast +}$, $\Gamma_{\rm tot} (D^{\ast +}) = 141$ keV is
consistent with the upper limit $\Gamma_{\rm tot} (D^{\ast +}) < 131$ keV
published by the ACCMOR Collaboration [11], more precision measurements of
the quantity are needed.

For the radiative decays $\Sigma_c \rightarrow \Lambda_c \gamma $ and
$\Xi^\prime_c \rightarrow \Xi_c \gamma$ the two light quarks in the initial
states have spin $1$, while they have spin $0$ in the final states.
Consequently, the diquark system must undergo a spin-flip transition.  The
charmed quark is a spectator in these transitions. Therefore, our
predictions for these decays are independent of the magnetic moment of the
charmed quark.

Both the chiral symmetry and the heavy quark symmetry play a critical role
in radiative decays involving pions.  The heavy quark symmetry relates the
strong coupling constants in the various pion emission vertices, while the
chiral symmetry dictates the structure of those vertices.  The specific
decays $\Sigma^0_c \rightarrow \Lambda^+_c \pi^- \gamma $ and $\Sigma^+_c
\rightarrow \Lambda^+_c \pi^0 \gamma $ are discussed in Section IV to
expose the essential features of these processes.

\bigskip\bigskip
\noindent{\bf II.~~Chiral Lagrangians for Electromagnetic Interactions of
Heavy Mesons}
\bigskip

To set up our notation, we denote the three light quarks by $q$

$$
q= \left( \begin{array}{c} u \\ d \\ s\\ \end{array}        \right)~,
\eqno(2.1)
$$

\noindent and the associated charge matrix by ${\cal{Q}} = {\rm diag}
(\frac{2}{3} , - \frac{1}{3} , - \frac{1}{3} )$.  The charge of the heavy
quark $Q$ is interchangeably denoted by $e_Q$ or
${\cal{Q}}^\prime$, depending on the circumstance of which one is more
convenient to use.  Under the electromagnetic gauge transformation

$$
A_\mu \rightarrow A^\prime_\mu = A_\mu - \frac{1}{e} \partial_\mu \lambda ,
\eqno(2.2)
$$

\noindent where $\lambda$ is a U(1) gauge parameter, the quark fields
transform as

$$
q \rightarrow q^\prime = e^{i {\cal {Q}} \lambda} q ~,~~~Q \rightarrow
Q^\prime = e^{i {\cal{Q}}^\prime \lambda} Q~. \eqno(2.3)
$$

\noindent Since the Goldstone-boson fields $M$ given by

$$
M = \left( \begin{array}{ccc}
\frac{\pi^0}{\sqrt{2}} + \frac{\eta}{\sqrt{6}} & \pi^+ & K^+ \\ \pi^- & -
\frac{\pi^0}{\sqrt{2}} + \frac{\eta}{\sqrt{6}} & K^0 \\ K^- &
\overline{K}^0 & - \sqrt{\frac{2}{3}} \eta
        \end{array} \right)  \eqno(2.4)
$$

\noindent are constructed from a light quark and an antiquark, they
transform as

$$
M \rightarrow M^\prime = e^{i {\cal {Q}} \lambda} M e^{-i {\cal{Q}}
\lambda} . \eqno(2.5)
$$

\noindent The meson field $\xi =$ exp $(\frac{iM}{\sqrt{2} f_\pi})$ thus
has a simple gauge transformation property

$$
\xi \rightarrow \xi^\prime = e^{i {\cal {Q}} \lambda} \xi e^{-i {\cal{Q}}
\lambda} ,~~\xi^\dagger \rightarrow \xi^{\prime \dagger} = e^{i {\cal{Q}}
\lambda} \xi^\dagger e^{-i {\cal{Q}} \lambda} .
\eqno(2.6)
$$

\noindent A gauge covariant derivative of the field $\xi$ has the form

$$
D_\mu \xi = \partial_\mu \xi + i e A_\mu [ {\cal {Q}}, ~\xi ] ,
\eqno(2.7)
$$

\noindent with the gauge transformation

$$
D_\mu \xi \rightarrow D^\prime_\mu \xi^\prime = e^{i {\cal{Q}} \lambda}
(D_\mu \xi) e^{-i {\cal{Q}} \lambda} .
\eqno(2.8)
$$

In the presence of electromagnetic interactions, the vector and axial
vector fields defined by

$$
{\cal{V}}^{(0)}_\mu = \frac{1}{2} (\xi^\dagger \partial_\mu \xi + \xi
\partial_\mu \xi^\dagger ), \eqno(2.9a)
$$

$$
{\cal{A}}^{(0)}_\mu = \frac{i}{2} (\xi^\dagger \partial_\mu \xi - \xi
\partial_\mu \xi^\dagger), \eqno(2.9b)
$$

\noindent become

$$
{\cal{V}}_\mu = \frac{1}{2}\,[ \xi^\dagger D_\mu \xi + \xi (D_\mu
\xi)^\dagger ] , \eqno(2.10a)
$$

$$
{\cal{A}}_\mu = \frac{i}{2}\,[ \xi^\dagger D_\mu \xi - \xi (D_\mu
\xi)^\dagger ] , \eqno(2.10b)
$$

\noindent where we have used the calligraphic letters ${\cal{V}}_\mu$ and
${\cal{A}}_\mu$ to denote the chiral vector and axial fields, respectively.
More explicitly, ${\cal{V}}_\mu$ and ${\cal{A}}_\mu$ are related to
${\cal{V}}^{(0)}_\mu$ and ${\cal{A}}^{(0)}_\mu$ respectively by

$$
{\cal{V}}_\mu = {\cal{V}}^{(0)}_\mu -i e {\cal{Q}} A_\mu + i \frac{1}{2} e
A_\mu (\xi^\dagger {\cal{Q}} \xi + \xi {\cal{Q}} \xi^\dagger ) ,
\eqno(2.11a)
$$

$$
{\cal{A}}_\mu = {\cal{A}}^{(0)}_\mu - \frac{1}{2} e A_\mu (\xi^\dagger
{\cal{Q}} \xi - \xi {\cal{Q}} \xi^\dagger ) ,
\eqno(2.11b)
$$

$$
{\cal{V}}^\ast_\mu = {\cal{V}}^{(0)\ast}_\mu + i e {\cal{Q}} A_\mu - i
\frac{1}{2} e A_\mu (\xi^T {\cal{Q}} \xi^\ast + \xi^\ast {\cal{Q}} \xi^T ) ,
\eqno(2.11c)
$$

$$
{\cal{A}}^\ast_\mu = {\cal{A}}^{(0)\ast}_\mu - \frac{1}{2} e A_\mu (\xi^T
{\cal{Q}} \xi^\ast - \xi^\ast {\cal{Q}} \xi^T ) ,
\eqno(2.11d)
$$

\noindent where we have given ${\cal{V}}^\ast_\mu$ and ${\cal{A}}^\ast_\mu$
since they appear in the following discussion.  The complex conjugate is
related to operation of hermitian conjugation and transposition, for
example, ${\cal{V}}^\ast_\mu = ({\cal{V}}^\dagger_\mu)^T$.

We next turn to the gauge transformation properties of heavy mesons.
Following the notation of Ref.[3], the ground-state $1^-$ and $0^-$ heavy
mesons are denoted by $P^\ast$ and $P$ respectively.  Since a heavy meson
contains a heavy quark $Q$ and a light antiquark $\overline{q}$, it obeys
the gauge transformation law

$$
P \rightarrow P^\prime = e^{i {\cal{Q}}^\prime \lambda} P e^{-i {\cal{Q}}
\lambda} , \eqno(2.12)
$$

\noindent and a similar equation for the vector meson $P^\ast$.  An
electromagnetic gauge covariant derivative can then be constructed to be

$$
D_\mu P = \partial_\mu P + i e A_\mu ({\cal{Q}}^\prime P - P {\cal{Q}}) ,
\eqno(2.13)
$$

\noindent which transforms as

$$
D_\mu P \rightarrow D^\prime_\mu P^\prime = e^{i {\cal{Q}}^\prime \lambda}
(D_\mu P) e^{-i {\cal{Q}} \lambda} . \eqno(2.14)
$$

\noindent When the chiral field is included, the covariant derivative
finally reads (see Ref.[3])

$$
\begin{array}{lcl}
D_\mu P & = & \partial_\mu P + {\cal{V}}^\ast_\mu P + i e A_\mu
({\cal{Q}}^\prime P - P {\cal{Q}}) , \\ ~~ \\
 & = & \partial_\mu P + {\cal{V}}^{(0) \ast}_\mu P + i e {\cal{Q}}^\prime
A_\mu P - i \frac{1}{2} e A_\mu (\xi^T {\cal{Q}} \xi^\ast + \xi^\ast
{\cal{Q}} \xi^T ) P , \end{array}  \eqno(2.15a)
$$

\noindent where use of Eq. (2.11c) has been made.  Similarly,

$$
\begin{array}{lcl}
D_\mu P^\dagger & = & \partial_\mu P^\dagger + {\cal{V}}_\mu P^\dagger - i
e A_\mu (P^\dagger {\cal{Q}}^\prime - {\cal{Q}} P^\dagger ) \\ ~~ \\
 & = & \partial_\mu P^\dagger + {\cal{V}}^{(0)}_\mu P^\dagger - i e
{\cal{Q}}^\prime A_\mu P^\dagger + \frac{i}{2} e A_\mu (\xi^\dagger
{\cal{Q}} \xi + \xi {\cal{Q}} \xi^\dagger ) P^\dagger . \end{array}
\eqno(2.15b)
$$

\noindent Eq. (2.15) shows that the electromagnetic interactions break the
SU(3) flavor symmetry.  The charge operator ${\cal{Q}}$ has an equal
mixture of $8_L$ and $8_R$ as it should be since the electromagnetic
interactions conserve parity.  The construction of the electromagnetic
gauge invariant chiral Lagrangian for heavy mesons simply follows from
gauging the chiral invariant meson Lagrangian presented in Ref.[3].  The
relevant terms are

$$
\begin{array}{lcl}
{\cal{L}}^{(1)}_{PP^\ast} & = & D_\mu P D^\mu P^\dagger -M^2_P P
P^\dagger + f \sqrt{M_P M_{P^\ast}} (P {\cal{A}}^\mu P^{\ast \dagger}_\mu +
P^\ast_\mu {\cal{A}}^\mu P^\dagger ) \\ ~~ \\
 &  & - \frac{1}{2} P^{\ast \mu \nu} P^{\ast \dagger}_{\mu \nu} +
M^2_{P^\ast} P^{\ast \mu} P^{\ast \dagger}_\mu + \frac{1}{4} f
\epsilon_{\mu \nu \lambda \kappa} (P^{\ast \mu \nu} {\cal{A}}^\lambda
P^{\ast \kappa \dagger} + P^{\ast \kappa} {\cal{A}}^\lambda P^{\ast \mu \nu
\dagger} ) ,
\end{array} \eqno(2.16)
$$

\noindent where Eqs. (2.11) and (2.15) have been used,

$$
P^{\ast \dagger}_{\mu \nu} = D_\mu P^{\ast \dagger}_\nu - D_\nu P^{\ast
\dagger}_\mu , \eqno(2.17)
$$

\noindent and $D_\mu P^{\ast \dagger}_\nu$ is given by Eq. (2.15b) with
$P^\dagger$ replaced by $P^{\ast \dagger}_\nu$.  The universal coupling
constant $f$ is independent of heavy quark masses and species.  By
expanding the meson field matrix $\xi$ into a power series

$$
\xi = 1 + i \frac{M}{\sqrt{2} f_\pi} - \frac{M^2}{4 f^2_\pi} + \cdots ,
\eqno(2.18)
$$

\noindent it is evident that ${\cal{V}}_\mu ({\cal{A}}_\mu)$ contains only
even (odd) number of pions interacting electromagnetically.  Consequently,
the kinematic terms in (2.16) give rise to contact terms with one photon
and even-number pion emissions, while the interacting terms yield
electromagnetic contact terms with odd-number Goldstone boson emission.

Note that the radiative transition $P^\ast \rightarrow P \gamma$ cannot
arise from the electromagnetic Lagrangian (2.16).  The lowest-order gauge
and chiral invariant interaction that contributes to $P^\ast \rightarrow P
\gamma$ is

$$
\begin{array}{ll}
{\cal{L}}^{(2)}_{PP^\ast} = & \sqrt{M_P M_{P^\ast}} \epsilon_{\mu \nu
\alpha \beta} v^\alpha P^{\ast \beta} \left[ \frac{1}{2}d ( \xi^\dagger
{\cal{Q}} \xi + \xi {\cal{Q}} \xi^\dagger) + d^\prime {\cal{Q}}^\prime
\right] F^{\mu \nu} P^\dagger + h.c. \\ ~~ \\
 & -ie F_{\mu \nu} P^{\ast \nu} \left[ {\cal{Q}}^\prime - \frac{1}{2} (
\xi^\dagger {\cal{Q}} \xi + \xi {\cal{Q}} \xi^\dagger) \right] P^{\ast \mu
\dagger} \\ ~~ \\
 & + i d^{\prime  \prime} M_{P^\ast} F_{\mu \nu} P^{\ast \nu} \left[ \gamma
{\cal{Q}}^\prime - \frac{1}{2} (\xi^\dagger {\cal{Q}} \xi + \xi {\cal{Q}}
\xi^\dagger) \right] P^{\ast \mu \dagger} . \end{array}
\eqno(2.19)
$$

\noindent In (2.19), $v^\alpha$ is the four-velocity of the $1^-$ heavy
meson and the second term is to remove the magnetic moment coupled to the
electromagnetic field implied by the minimal couplings in (2.16), while the
last term proportional to $d^{\prime \prime}$ is to account for the
magnetic moment couplings due to the constituent quarks, both light and
heavy.  The universal coupling constant $d$ is independent of the heavy
quark masses and species. We have also included the $d^\prime$ and $\gamma$
terms to account for the corrections due to the heavy quark masses when
$m_Q \neq \infty$.

The Lagrangian (2.19) describe the magnetic transitions $P^\ast \rightarrow
P \gamma $ and $P^\ast \rightarrow P^\ast \gamma $.  In the infinitely
heavy quark mass limit, only the two parameters $d$ and $d^{\prime \prime}$
in Eq. (2.19) survive.  The heavy quark spin symmetry then relates them.
To derive the relation, we will make use of the interpolating fields
introduced in Ref.[3]

$$
\begin{array}{rcl}
P(v) & = & \overline{q}_v  \gamma_5 h_v \sqrt{M_P} , \\ ~~ \\
P^\ast (v, \epsilon) & = & \overline{q}_v \not{\epsilon} h_v
\sqrt{M^\ast_P} ,
\end{array} \eqno(2.20)
$$

\noindent where $\overline{q}_v$ is a light antiquark which combines with a
heavy quark $h_v$ of velocity $v$ to form the appropriate meson.  Now, let
$J_\mu$ and $j_\mu$ be the electromagnetic currents of the heavy quark and
light quarks, respectively.  It is easy to show that $J_\mu$ does not
contribute to the magnetic transitions of interest here.  Consider

$$
\begin{array}{lcl}
<P(v^\prime) \mid J_\mu \mid P^\ast (v, \epsilon) >  & = & \sqrt{M_P
M^\ast_P} <0 \mid \overline{q}_{v^\prime}  \gamma_5 h_{v^\prime}
\overline{h}_{v^\prime} \gamma_\mu h_v \overline{h}_v \not{\epsilon} q_v \mid
0 > \\ ~~ \\
 & = & - \sqrt{M_P M^\ast_P} {\rm tr} \left\{  \gamma_5
\frac{\not{v}^\prime + 1}{2} \gamma_\mu \frac{\not{v} + 1}{2}
\not{\epsilon} <0 \mid q_v \overline{q}_{v^\prime} \mid 0> \right\} .
\end{array}  \eqno(2.21)
$$

\noindent Sandwiched between the projection matrices, the matrix
$\gamma_\mu$ can be replaced by

$$
\gamma_\mu = \frac{1}{2} (v_\mu + v^\prime_\mu ) - \frac{1}{2m_Q}
\sigma_{\mu \nu} k^\nu , \eqno(2.22)
$$

\noindent which shows that in the limit $m_Q \rightarrow \infty$, the heavy
quark's electromagnetic current does not induce a magnetic coupling.  We
also notice that the heavy quark current is conserved by itself, so the
light quark current must be separately conserved.  We are now ready to
examine the electromagnetic vertices associated with the light quark
current.  We have

$$
<P(v^\prime) \mid j_\mu \mid P^\ast (v, \epsilon) > = - \sqrt{M_P
M_{P^\ast}} {\rm tr} \left\{  \gamma_5 \frac{\not{v} + 1}{2}
\not{\epsilon} L_\mu \right\} , \eqno(2.23)
$$

\noindent where

$$
L_\mu = < 0 \mid q_v j_\mu \overline{q}_{v^\prime} \mid 0 > . \eqno(2.24)
$$

\noindent Lorentz covariance implies

$$
L_\mu = c_1 (v+v^\prime)_\mu + c_2 \gamma_\mu + c_3 \sigma_{\mu \nu}
k^\nu . \eqno(2.25)
$$

\noindent Taking the trace, we find

$$
<P(v^\prime) \mid j_\mu \mid P^\ast (v, \epsilon) > = - 2c_3 \sqrt{M_P
M_{P^\ast}} \epsilon_{\mu \nu \alpha \beta} k^\nu v^\alpha \epsilon^\beta .
\eqno(2.26)
$$

\noindent Similarly, we have

$$
<P^\ast (v^\prime , \epsilon_f ) \mid j_\mu \mid P^\ast (v, \epsilon_i ) >
= - M_{P^\ast} {\rm tr} \left\{ \not{\epsilon}_f \frac{\not{v} + 1}{2}
\not{\epsilon}_i L_\mu \right\} .  \eqno(2.27)
$$

\noindent In taking the trace, the $c_2$ term does not contribute as a
result of $\epsilon_i \cdot v = 0~,~~{\rm and}~ \epsilon_f \cdot v =
\frac{k}{M_{P^\ast}} \approx 0$, while the $c_1$
term contributes, it is not of the magnetic type.  Thus

$$
<P^\ast (v^\prime , \epsilon_f ) \mid j_\mu \mid P^\ast (v, \epsilon_i) >_m
= 2ic_3 M_{P^\ast} (\epsilon_f \cdot k \epsilon_{i \mu} - \epsilon_i \cdot
k \epsilon_{f \mu} ) , \eqno(2.28)
$$

\noindent where the subscript $m$ is a reminder that we keep only the part
dependent on the magnetic moment.  By comparison with the matrix elements
implied by Eq. (2.19) for $P^\ast \rightarrow P \gamma $ and $P^\ast
\rightarrow P^\ast \gamma$, we find

$$
\begin{array}{lcl}
d & = & -ic_3 \\
d^{\prime \prime} & = & 2ic_3 ~~.
\end{array}  \eqno(2.29)
$$

\noindent The relation we are looking for is

$$
d^{\prime \prime} = -2d ~.  \eqno(2.30)
$$

The SU(3) breaking effects due to the light quark mass differences can be
incorporated in the Lagrangian (2.19) by replacing the charge matrix
${\cal{Q}}$ by (see also Sec. 3)

$$
{\cal{Q}} \rightarrow \stackrel{\sim}{\cal{Q}} = \left(
\begin{array}{ccc}
\frac{2}{3} & 0 & 0 \\
0 & - \frac{\alpha}{3} & 0 \\
0 & 0 & - \frac{\beta}{3}
\end{array} \right) ,  \eqno(2.31)
$$

\noindent where $\alpha = m_u/m_d$ and $\beta = m_u/m_s$.

We now show that the nonrelativistic quark model has a simple prediction
for the couplings $d$, $d^\prime$, $d^{\prime \prime}$ and $\gamma$.  The
magnetic interaction of the quarks is

$$
{\cal{L}}_{\rm em} = \sum_i e
\frac{e_i}{2m_i} \overline{\psi} \stackrel{\rightharpoonup}{\sigma_i} \psi
\cdot \stackrel{\rightharpoonup}{H} = \overline{\psi} ( \sum_i \mu_i
\stackrel{\rightharpoonup}{\sigma_i} )
\psi \cdot \stackrel{\rightharpoonup}{H} , \eqno(2.32)
$$

\noindent where $e_i$ is the charge of the $i$th quark in units of $e$.
Instead of using the Lorentz invariant normalization

$$
<P \mid P^\prime > = 2E (2 \pi)^3  \delta^3 ( P - P^\prime ) ,
\eqno(2.33)
$$

\noindent it is more convenient to use a discrete normalization by
enclosing the system in a large volume $V$, so that

$$
\ll P \mid P^\prime \gg = \delta_{\stackrel{\rightharpoonup}{P}
\stackrel{\rightharpoonup}{P^\prime}} , \eqno(2.34)
$$

\noindent Then in the rest frame we get

$$
<P \mid {\cal{L}}_{\rm em} \mid P^\ast > = 2 \sqrt{M_P M_{P^\ast}} \ll P
\mid \sum_q \mu_q \sigma^z_q - \sum_{\overline{q}} \mu_q
\sigma^z_{\overline{q}} \mid P^\ast \gg H , \eqno(2.35)
$$

\noindent where we have chosen the magnetic field along the $z$ direction,
and the minus sign for  the antiquarks can be understood simply as they
have charges opposite to those of quarks.  Next we need the flavor-spin
wave functions of the heavy mesons in the nonrelativistic quark model:

$$
\begin{array}{lcl}
\mid P^\ast \gg & = & \frac{1}{\sqrt{2}} \mid Q \uparrow
\overline{q} \downarrow
+ Q \downarrow \overline{q} \uparrow > , \\ ~~ \\
\mid P \gg & = & \frac{1}{\sqrt{2}} \mid Q \uparrow \overline{q} \downarrow
- Q \downarrow \overline{q} \uparrow > , \end{array}  \eqno(2.36)
$$

\noindent where $\mid P^\ast \gg$ denotes the vector meson state with the
$z$-component of its spin being zero.  Let us denote the SU(3) $P_i$ as

$$
P_i = (Q \overline{d} ,~ Q \overline{u} ,~ Q \overline{s} ) =
(P^{\frac{1}{2}} ,~ P^{-\frac{1}{2}},~ P^0 ) , \eqno(2.37)
$$

\noindent where the superscript indicates the isospin quantum number $I_3$.
We then  have

$$
\begin{array}{lcl}
<P^{\frac{1}{2}} \mid {\cal{L}}_{\rm em} \mid P^{\ast \frac{1}{2}} > & = &
2 \sqrt{M_P M_{P^\ast}} (\mu_d + \mu_Q) , \\ ~~ \\
<P^{-\frac{1}{2}} \mid {\cal{L}}_{\rm em} \mid P^{\ast -\frac{1}{2}} > & =
& 2 \sqrt{M_P M_{P^\ast}} (\mu_u + \mu_Q ) , \\ ~~ \\
<P^0 \mid {\cal{L}}_{\rm em} \mid P^{\ast 0} > & = & 2 \sqrt{M_P
M_{P^\ast}} (\mu_s + \mu_Q ) , \end{array}  \eqno(2.38)
$$

\noindent where we have dropped the magnetic field for convenience.

Note that in the rest frame of $P^\ast,~v^\alpha = (1,
\stackrel{\rightharpoonup}{0})$ so that

$$
\epsilon_{\mu \nu \alpha \beta} F^{\mu \nu} v^\alpha \epsilon^{\ast \beta}
= \epsilon_{ijk} F^{ij} \epsilon^{\ast k} = -2
\stackrel{\rightharpoonup}{\epsilon^\ast} \cdot
\stackrel{\rightharpoonup}{H} .
\eqno(2.39)
$$

Choosing the $\stackrel{\rightharpoonup}{H}$ field along the $z$ direction
as before, we find from (2.19), (2.31) and (2.39) that

$$
\begin{array}{lcl}
<P^{\frac{1}{2}} \mid {\cal{L}}^{(2)}_{PP^\ast} \mid P^{\ast \frac{1}{2}} >
& = & -2 \sqrt{M_P M_{P^\ast}} (-\frac{\alpha}{3} d + e_Q d^\prime)\\ ~~ \\
<P^{-\frac{1}{2}} \mid {\cal{L}}^{(2)}_{P P^\ast} \mid P^{\ast -
\frac{1}{2}} > & = & -2 \sqrt{M_P
M_{P^\ast}} ( \frac{2}{3} d + e_Q d^\prime ) , \\ ~~ \\
<P^0 \mid {\cal{L}}^{(2)}_{PP^\ast} \mid P^{\ast 0} > & = & -2 \sqrt{M_P
M_{P^\ast}} ( -\frac{\beta}{3} d + e_Q d^\prime ) , \end{array} \eqno(2.40)
$$

\noindent Comparing this with Eq. (2.38) gives the desired results

$$
d = -\frac{e}{2m_u} ~~,~~ d^\prime = - \frac{e}{2m_Q} . \eqno(2.41)
$$

\noindent A similar calculation gives

$$
d^{\prime \prime} = \frac{e}{m_u} ~~,~~ \gamma = \frac{m_u}{m_Q} .
\eqno(2.42)
$$

\noindent The quark model results (2.41) and (2.42) satisfy the heavy quark
symmetry relation (2.30).  This is not surprising, as SU(3) breakings
preserve the heavy quark symmetry.

\bigskip\bigskip
\noindent{\bf III.~~Chiral Lagrangians for Electromagnetic Interactions of
Heavy Baryons}
\bigskip

We consider a heavy baryon containing a heavy quark and two light quarks.
The two light quarks form either a symmetric sextet {\bf 6} or an
antisymmetric antitriplet {\bf \={3}} in flavor SU(3) space.  We will
denote these spin $\frac{1}{2}$ baryons as $B_6$ and $B_{\overline{3}}$
respectively, and the spin $\frac{3}{2}$ baryon by $B^\ast_6$.  Explicitly,
the baryon matrices read as in Ref.[3]

$$
B_6 = \left( \begin{array}{ccc}
\Sigma^{+1}_Q & \frac{1}{\sqrt{2}} \Sigma^0_Q & \frac{1}{\sqrt{2}}
\Xi^{\prime + \frac{1}{2}}_Q \\ ~~ \\
\frac{1}{\sqrt{2}} \Sigma^0_Q & \Sigma^{-1}_Q & \frac{1}{\sqrt{2}}
\Xi^{\prime - \frac{1}{2}}_Q \\ ~~ \\
\frac{1}{\sqrt{2}} \Xi^{\prime + \frac{1}{2}}_Q & \frac{1}{\sqrt{2}}
\Xi^{\prime - \frac{1}{2}}_Q & \Omega_Q
 \end{array}  \right) ,
\eqno(3.1)
$$

$$
B_{\overline{3}} = \left( \begin{array}{ccc}
0 & \Lambda_Q & \Xi^{+ \frac{1}{2}}_Q \\ ~~ \\
-\Lambda_Q & 0 & \Xi^{- \frac{1}{2}}_Q \\ ~~ \\
-\Xi^{+ \frac{1}{2}}_Q & - \Xi^{- \frac{1}{2}}_Q & 0
  \end{array} \right) ,
\eqno(3.2)
$$

\noindent and a matrix for $B^\ast_6$ similar to $B_6$.  The
superscript in (3.1) and (3.2) refers to the value of the isospin quantum
number $I_3$.

Under the electromagnetic gauge transformation Eq. (2.2), the heavy baryon
field transforms as

$$
B \rightarrow B^\prime = e^{i {\cal{Q}}^\prime \lambda} e^{i {\cal{Q}}
\lambda} B e^{i {\cal{Q}} \lambda} .
\eqno(3.3)
$$

\noindent It is then easily shown that the electromagnetic gauge covariant
derivative has the form

$$
D_\mu B = (\partial_\mu + i e {\cal{Q}}^\prime A_\mu) B + i e A_\mu \left\{
{\cal{Q}},~ B \right\} ,
\eqno(3.4)
$$

\noindent which transforms according to

$$
D_\mu B \rightarrow D^\prime_\mu B^\prime = e^{i {\cal{Q}}^\prime \lambda}
e^{i {\cal{Q}} \lambda} (D_\mu B) e^{i {\cal{Q}} \lambda} .
\eqno(3.5)
$$

\noindent With the chiral fields included, the covariant derivative is
modified to (see Ref.[3])

$$
D_\mu B = \partial_\mu B + {\cal{V}}_\mu B + B {\cal{V}}^T_\mu + i e
{\cal{Q}}^\prime A_\mu B + i e A_\mu \left\{ {\cal{Q}} ,~ B \right\} .
\eqno(3.6)
$$

\noindent It follows from Eq. (2.11a) that

$$
\begin{array}{c}
D_\mu B = \partial_\mu B + {\cal{V}}^{(0)}_\mu B + B {\cal{V}}^{(0) T}_\mu
+ i e {\cal{Q}}^\prime A_\mu B \\ ~~ \\
+ i \frac{1}{2} e A_\mu \left[ (\xi^\dagger {\cal{Q}} \xi + \xi {\cal{Q}}
\xi^\dagger ) B + B (\xi^\dagger {\cal{Q}} \xi + \xi {\cal{Q}} \xi^\dagger
)^T      \right] .
   \end{array}  \eqno(3.7)
$$

As in the meson case discussed in the previous section, a chiral and
electromagnetic gauge invariant Lagrangian for heavy baryons can be
obtained by gauging the chiral Lagrangian (3.12) given in Ref.[3].  We
write down the relevant terms

$$
\begin{array}{lcl}
{\cal{L}}^{(1)}_B & = & \frac{1}{2} \tr [\overline{B}_{\overline{3}} (i
{\not{D}} - M_{\overline{3}}) B_{\overline{3}} ] + \tr [ \overline{B}_6 (i
{\not{D}} - M_6) B_6 ] \\ ~~ \\
& + & \tr \left\{ \overline{B}^{\ast \mu}_6 [ -g_{\mu \nu} ( i {\not{D}} -
M_{6^\ast}) + i (\gamma_\mu D_\nu + \gamma_\nu D_\mu ) - \gamma_\mu (i
{\not{D}} + M_{6^\ast}) \gamma_\nu ] B^{\ast \nu}_6 \right\} \\ ~~ \\
 & + & g_1 \tr ( \overline{B}_6 \gamma_\mu \gamma_5 {\cal{A}}^\mu B_6 ) +
g_2 \tr (\overline{B}_6 \gamma_\mu \gamma_5 {\cal{A}}^\mu B_{\overline{3}} )
+ h.c. \\ ~~ \\
 & + & \frac{\sqrt{3}}{2} g_1 \tr (\overline{B}^\ast_{6 \mu} {\cal{A}}^\mu
B_6 ) + h.c. - \sqrt{3} g_2 \tr ( \overline{B}^{\ast \mu}_6 {\cal{A}}_\mu
B_{\overline{3}} ) + h.c. \\ ~~ \\
 & - & \frac{3}{2} g_1 \tr ( \overline{B}^{\ast \nu}_6 \gamma_\mu \gamma_5
{\cal{A}}^\mu B^\ast_{6 \nu} ) , \end{array}
\eqno(3.8)
$$

\noindent with $D_\mu B$ and ${\cal{A}}_\mu$ given by (3.7)
and (2.11), respectively, where $B^\ast_{6 \mu}$ is a Rarita-Schwinger's
vector-spinor field for  a spin $\frac{3}{2}$ particle, and use of heavy
quark symmetry has been applied to relate various coupling constants.  As
in the case of heavy mesons, electromagnetic contact terms with even (odd)
number of pions come from kinematic (interacting) terms in (3.8).

Since baryons do not behave much like Dirac point particles, they can have
large anomalous magnetic moments.  Apart from the non-anomalous
electromagnetic interaction described by ${\cal{L}}^{(1)}_B$, the most
general electromagnetic invariant Lagrangian for  anomalous magnetic
transitions of heavy baryons is given by (we use the abbreviation $\sigma
\cdot F \equiv \sigma_{\mu \nu} F^{\mu \nu}$.)

$$
\begin{array}{lcl}
{\cal{L}}^{(2)}_B & = & a_1 {\rm tr} (\overline{B}_6 {\cal{Q}} \sigma \cdot
F B_6 ) + a^\prime_1 {\rm tr} (\overline{B}_6 {\cal{Q}}^\prime \sigma \cdot
F B_6 ) \\ ~~ \\
& + & a_2 {\rm tr} (\overline{B}_6 {\cal{Q}} \sigma \cdot F
B_{\overline{3}} ) + h.c. + a^\prime_2 {\rm tr} (\overline{B}_6
{\cal{Q}}^\prime \sigma \cdot F B_{\overline{3}} ) + h.c. \\ ~~ \\
& + & a_3 {\rm tr} (\epsilon_{\mu \nu \lambda \kappa} \overline{B}^{\ast
\mu}_6 {\cal{Q}} \gamma^\nu F^{\lambda \kappa} B_6 ) + h.c. + a^\prime_3
{\rm tr} (\epsilon_{\mu \nu \lambda \kappa} \overline{B}^{\ast \mu}_6
{\cal{Q}}^\prime \gamma^\nu F^{\lambda \kappa} B_6 ) + h.c. \\ ~~ \\
& + & a_4 {\rm tr} (\epsilon_{\mu \nu \lambda \kappa} \overline{B}^{\ast
\mu}_6 {\cal{Q}} \gamma^\nu F^{\lambda \kappa} B_{\overline{3}} ) + h.c. +
a^\prime_4 {\rm tr} (\epsilon_{\mu \nu \lambda \kappa} \overline{B}^{\ast
\mu}_6 {\cal{Q}}^\prime \gamma^\nu F^{\lambda \kappa} B_{\overline{3}} ) +
h.c. \\ ~~ \\
& + & a_5 {\rm tr} ( \overline{B}^{\ast \mu}_6 {\cal{Q}} \sigma \cdot F
B^\ast_{6 \mu} ) + a^\prime_5 {\rm tr} (\overline{B}^{\ast \mu}_6
{\cal{Q}}^\prime \sigma \cdot F B^\ast_{6 \mu} ) \\ ~~ \\
& + & a_6 {\rm tr} ( \overline{B}_{\overline{3}} {\cal{Q}} \sigma \cdot F
B_{\overline{3}} ) + a^\prime_6 {\rm tr} ( \overline{B}_{\overline{3}}
{\cal{Q}}^\prime \sigma \cdot F B_{\overline{3}} ) \\ ~~ \\
& + & \frac{1}{4} \mu_{_{B_{\overline{3}}}} {\rm tr} (
\overline{B}_{\overline{3}} {\cal{Q}}_{\rm tot} \sigma \cdot F
B_{\overline{3}} ) + \frac{1}{2} \mu_{_{B_6}} {\rm tr} ( \overline{B}_6
{\cal{Q}}_{\rm tot} \sigma \cdot F B_6) \\ ~~ \\
& - & \frac{1}{2} \mu_{_{B^\ast_6}} {\rm tr} (\overline{B}^{\ast \mu}_6
{\cal{Q}}_{\rm tot} \sigma \cdot F B^\ast_{6 \mu} ) ,
  \end{array} \eqno(3.9)
$$

\noindent where ${\cal{Q}}_{\rm tot} = 2{\cal{Q}} + {\cal{Q}}^\prime$, and
$\mu_B = \frac{e}{2M_B}$, recalling that ${\cal{Q}}$ is the charge matrix
of light quarks, whereas ${\cal{Q}}^\prime$ (or $e_Q$) is the charge of the
heavy quark.  The Lagrangian ${\cal{L}}^{(2)}_B$ is also the most general
chiral-invariant one provided that one makes the replacement

$$
{\cal{Q}} \rightarrow \frac{1}{2} ( \xi^\dagger {\cal{Q}} \xi + \xi
{\cal{Q}} \xi^\dagger ) ~,~ {\cal{Q}}^\prime \rightarrow {\cal{Q}}^\prime .
\eqno(3.10)
$$

\noindent Note that we have substracted out the Dirac magnetic moments of
heavy baryons $\mu_B$, so that in the quark model the anomalous magnetic
moments $a_i$ are simply related to the Dirac magnetic moments of the light
quarks, while $a^\prime_i$ are connected to those of the heavy quarks.
In the heavy quark limit, both Dirac magnetic moments $\mu_B$ and the
heavy-quark magnetic moments $a^\prime_i$ vanish as they are suppressed by
the heavy quark mass.

At first sight, it appears that other gauge invariants e.g.,
$\overline{B}^{\ast \mu}_6 F_{\mu \nu} \gamma^\nu \gamma_5 B_6$, and
$\overline{B}^{\ast \mu}_6 F_{\mu \nu} B^{\ast \nu}_6$ can be added to
${\cal{L}}^{(2)}_B$.  However, by applying the identity

$$
i \epsilon^{\mu \nu \lambda \kappa} \gamma_\kappa = - \gamma^\mu \gamma^\nu
\gamma^\lambda \gamma_5 + g^{\mu \nu} \gamma^\lambda \gamma_5 - g^{\mu
\lambda} \gamma^\nu \gamma_5 + g^{\nu \lambda} \gamma^\mu \gamma_5 ,
\eqno(3.11)
$$

\noindent we see that

$$
{\rm tr} (\overline{B}^{\ast \mu}_6 F_{\mu \nu} \gamma^\nu \gamma_5 B) =
\frac{i}{2} {\rm tr} (\epsilon_{\mu \nu \lambda \kappa}
\overline{B}^{\ast \mu}_6 \gamma^\nu F^{\lambda \kappa} B) ,
\eqno(3.12)
$$

\noindent for $B=B_{\overline{3}}$ or $B_6$.  Next, using the fact that the
Rarita-Schwinger vector spinor $u_\lambda$ obeys the relations

$$
u_\mu = i \sigma_{\mu \nu} u^\nu,~~~~~~\overline{u}_\mu = i
\overline{u}^\nu \sigma_{\nu \mu} ,
\eqno(3.13)
$$

\noindent it is straightforward to show that

$$
\begin{array}{lcl}
\overline{u}^\mu F_{\mu \nu} u^\nu & = & i^2\, \overline{u}^\lambda
\sigma_\lambda^{~~\mu} F_{\mu \nu} \sigma^\nu_{~\kappa} u^\kappa \\ ~~ \\
& = & \overline{u}^\mu F_{\mu \nu} u^\nu + 2 \overline{u}^\nu F_{\mu \nu}
u^\mu + i \overline{u}^\lambda \epsilon_{\lambda \mu \nu \kappa} \gamma_5
F^{\mu \nu} u^\kappa \\ ~~ \\
& & - i \overline{u}^\lambda \sigma_{\mu \nu} F^{\mu \nu} g_{\lambda
\kappa} u^\kappa ~~,   \end{array} \eqno(3.14)
$$

\noindent and hence

$$
\overline{u}^\lambda (v^\prime) \sigma \cdot F u_\lambda (v) = 2 i
\overline{u}^\mu (v^\prime) F_{\mu \nu} u^\nu (v) + \overline{u}^\lambda
(v^\prime) \epsilon_{\lambda \mu \nu \kappa} \gamma_5 F^{\mu \nu} u^\kappa
(v).  \eqno(3.15)
$$

\noindent In the heavy quark limit and $v^\prime \sim v$,
$\overline{u}^\lambda (v) \gamma_5 u^\kappa (v) = 0$.  Therefore, there are
only six independent couplings in the heavy quark limit for anomalous
magnetic moment radiative baryonic transitions.

We shall see that the heavy quark spin symmetry reduces the six couplings
$a_i$ to two independent ones.  To embark on this task, we will apply the
interpolating fields for the heavy baryons in terms of the diquark fields
of the light quarks (see Ref.[3])

$$
B_{\overline{3}} (v,s) = \overline{u} (v,s) \phi_v h_v ,
\eqno(3.16)
$$

$$
B_6 (v,s,\kappa) = \overline{B}_\mu (v,s,\kappa) \phi^\mu_v h_v ,
\eqno(3.17)
$$

\noindent where $\phi_v$ and $\phi^\mu_v$ are the $0^+$ and $1^+$ diquarks,
respectively, which combine with the heavy quark $h_v$ of velocity $v$ to
form the appropriate heavy baryon.  The argument $\kappa$ indicates the
spin of the baryon: $\kappa = 1$ for spin $\frac{1}{2}$ baryons $(B_6)$ and
$\kappa=2$ for spin $\frac{3}{2}$ baryons $(B^\ast_6)$.  The wave function
$\overline{B}_\mu$ is given by

$$
\overline{B}_\mu (v,s, \kappa = 1) = \frac{1}{\sqrt{3}} \overline{u} (v,s)
\gamma_5 (v_\mu + \gamma_\mu) ,
\eqno(3.18a)
$$

$$
\overline{B}_\mu (v,s, \kappa = 2) = \overline{u}_\mu (v,s) .
\eqno(3.18b)
$$

We shall now apply heavy quark symmetry to the magnetic-moment coupling
constants $a_i$.  As in the meson case, let us denote the electromagnetic
current of light and heavy quarks by $j_\mu$ and $J_\mu$, respectively.
For the couplings $a_1 \cdots a_6$, we do not have to consider the heavy
quark current.  For example, it is easily shown that

$$
<B_{\overline{3}} (v^\prime , s^\prime) \mid J_\mu \mid B_{\overline{3}}
(v,s) > = e_Q \zeta (v \cdot v^\prime) \overline{u} (v^\prime, s^\prime)
\gamma_\mu u (v,s) , \eqno(3.19)
$$

\noindent where $\zeta (v \cdot v^\prime) = < 0 \mid \phi_{v^\prime}
\phi^\dagger_{v} \mid 0 >$ is a universal Isgur-Wise function.  Eq. (2.22)
is applicable here, and it shows that the heavy quark electromagnetic
current does not induce a magnetic-type coupling in the heavy quark limit.
As in the meson case, the heavy quark current is conserved by itself, so
the light quark current must be separately conserved. We next note that
$a_6 = 0 $ because the spin of the heavy quark cannot be flipped by a
photon emission and because the radiative transition $0^+ \rightarrow 0^+ +
\gamma$ in the diquark sector is prohibited by conservation of angular
momentum.  Indeed, the interpolating field method gives

$$
\begin{array}{l}
<B_{\overline{3}} (v^\prime , s^\prime) \mid j_\mu \mid B_{\overline{3}}
(v, s) > \\ ~~ \\
= < 0 \mid \overline{u} (v^\prime , s^\prime) \phi_{v^\prime} h_{v^\prime}
j_\mu \overline{h}_v \phi^\dagger_v u (v,s) \mid 0 > \\ ~~ \\
= \overline{u} (v^\prime , s^\prime) M_\mu u (v,s) ~~,
  \end{array}  \eqno(3.20)
$$

\noindent with

$$
M_\mu = < 0 \mid \phi_{v^\prime} j_\mu \phi^\dagger_v \mid 0 > .
\eqno(3.21)
$$

\noindent Now Lorentz invariance implies that

$$
M_\mu = a (v+v^\prime)_\mu + b k_\mu ,
\eqno(3.22)
$$

\noindent Since $k^\mu (v+v^\prime)_\mu = 0$, it is clear that conservation
of the electromagnetic current implies $b=0$.  Consequently,

$$
<B_{\overline{3}} (v^\prime , s^\prime) \mid j_\mu \mid B_{\overline{3}}
(v,s) > = a \overline{u} (v^\prime , s^\prime)(v+v^\prime)_\mu u(v,s) ,
\eqno(3.23)
$$

\noindent which is nothing but the usual convection current due to the
charge.  We thus conclude that

$$
a_6 =0 , \eqno(3.24)
$$

\noindent in the heavy quark limit.

We now turn to the matrix element of the $B_6 - B_6$ transition.  We have

$$
<B_6 (v^\prime , s^\prime) \mid j_\mu \mid B_6 (v, s) > =
\overline{B}^\alpha (v^\prime , s^\prime) M_{\mu \alpha \beta} B^{\beta} (v,
s) , \eqno(3.25)
$$

\noindent where

$$
M^{\mu \alpha \beta} = < 0 \mid \phi_{v^\prime}^\alpha j^\mu \phi^{\beta
\dagger}_v \mid 0 > . \eqno(3.26)
$$

\noindent The general expression of $M_{\mu \alpha \beta}$ linear in $k$ is

$$
M_{\mu \alpha \beta} = f_1 g_{\alpha \beta} (v+v^\prime)_\mu + f_2
g_{\alpha \beta} k_\mu + f_3 g_{\mu \alpha} k_\beta + f_4 g_{\mu \beta}
k_\alpha + f_5 v_\alpha v^\prime_\beta (v+v^\prime)_\mu ~~,
\eqno(3.27)
$$

\noindent where we do not display form factors proportional to
$v^\prime_\alpha$ or $v_\beta$ because of $\overline{B}_\alpha v^{\prime
\alpha} = 0$ and $v^\beta B_\beta = 0$.  Conservation of the
electromagnetic vector current then indicates that

$$
f_2 = 0~~,~~f_3 + f_4 = 0 , \eqno(3.28)
$$

\noindent and hence

$$
M_{\mu \alpha \beta} = f_1 g_{\alpha \beta} (v+v^\prime)_\mu + f_3 (g_{\mu
\alpha} k_\beta - g_{\mu \beta} k_\alpha) + f_5 v_\alpha v^\prime_\beta
(v+v^\prime)_\mu . \eqno(3.29)
$$

\noindent Using the interpolating field (3.17) we find

$$
\begin{array}{cl}
< & B_6 (v^\prime , s^\prime) \mid j_\mu \mid B_6 (v,s) > \\ ~~ \\
= & - \frac{1}{3} \overline{u} (v^\prime , s^\prime) \gamma_5 (\gamma^\alpha
+ v^{\prime \alpha}) M_{\mu \alpha \beta} (\gamma^\beta + v^\beta ) u (v,s)
\\ ~~ \\
= & - \frac{1}{3} \overline{u} (v^\prime , s^\prime) \bigg\{ \bigg( f_1 (2+v
\cdot v^\prime ) - f_5 [1-(v \cdot v^\prime)^2] \bigg) (v+v^\prime)_\mu
  \\~~ \\
& + f_3 (\gamma_\mu {\not{k}} - {\not{k}} \gamma_\mu ) \bigg\} u
(v,s)~~. \end{array}
\eqno(3.30)
$$

\noindent This leads to

$$
<B_6 (v^\prime , s^\prime) \gamma (k, \epsilon) \mid j_\mu A^\mu \mid B_6
(v, s)> = -\frac{1}{3} f_3 \overline{u} (v^\prime , s^\prime) \sigma \cdot
F u (v, s) .
\eqno(3.31)
$$

\noindent with $\sigma \cdot F = \sigma_{\mu \nu} F^{\mu \nu}~,~~~F^{\mu
\nu} = i (k^\mu \epsilon^\nu - k^\nu \epsilon^\mu )$ and

$$
a_1 = -\frac{1}{3} f_3 , \eqno(3.32)
$$

\noindent where we have dropped a convection current term.  Likewise, for
the magnetic $B^\ast_6 - B_6$ coupling, we get

$$
\begin{array}{lcl}
<B^\ast_6 (v^\prime , s^\prime) \mid j_\mu \mid B_6 (v, s) > & = &
-\frac{1}{\sqrt{3}} \overline{u}^\alpha (v^\prime , s^\prime) M_{\mu \alpha
\beta} (v^\beta + \gamma^\beta) \gamma_5 u (v, s), \\ ~~ \\
 & = & -\frac{1}{\sqrt{3}} f_3 \overline{u}^\alpha (v^\prime , s^\prime)
(g_{\mu \alpha}\not{k} - k_\alpha \gamma_\mu) \gamma_5 u (v, s) ,
  \end{array}    \eqno(3.33)
$$

\noindent and hence

$$
<B^\ast_6 (v^\prime, s^\prime) \gamma (k, \epsilon) \mid j_\mu A^\mu \mid
B_6 (v, s) > = -i \frac{f_3}{\sqrt{3}} \overline{u}^\mu (v^\prime ,
s^\prime) \gamma^\nu \gamma_5 F_{\mu \nu} u (v,s) , \eqno(3.34)
$$

\noindent where only the magnetic-type terms contribute.  Comparing this
with (3.9) and applying the relation (3.12) yields

$$
a_3 = \frac{1}{2 \sqrt{3}} f_3 . \eqno(3.35)
$$
Similarly,

$$
\begin{array}{lcl}
<B^\ast_6 (v^\prime , s^\prime) \mid j_\mu \mid B^\ast_6 (v,s) > & = &
\overline{u}^\alpha (v^\prime , s^\prime) M_{\mu \alpha \beta} u^\beta
(v,s) \\ ~~ \\
  & = & f_3 \overline{u}^\alpha (v^\prime , s^\prime) (g_{\mu \alpha}
k_\beta - g_{\mu \beta} k_\alpha ) u^\beta (v, s) , \end{array}
 \eqno(3.36)
$$

\noindent and

$$
<B^\ast_6 (v^\prime , s^\prime) \gamma (k, \epsilon) \mid j_\mu A^\mu \mid
B^\ast_6 (v,s) > = i f_3 \overline{u}^\alpha (v^\prime , s^\prime)
F_{\alpha \beta} u^\beta (v,s) . \eqno(3.37)
$$

\noindent This together with Eq. (3.15) leads to

$$
a_5 = \frac{f_3}{2}. \eqno(3.38)
$$

\noindent It follows from Eqs. (3.32), (3.35) and (3.38) that the coupling
constants $a_1, a_3$ and $a_5$ are related via heavy quark symmetry.

We next turn to the $a_2$ term and get

$$
<B_6 (v^\prime , s^\prime) \mid j_\mu \mid B_{\overline{3}} (v,s) > =
\overline{u}^\nu (v^\prime , s^\prime) M_{\mu \nu} u(v,s) , \eqno(3.39)
$$

\noindent with

$$
M^{\mu \nu} = < 0 \mid \phi^\nu_{v^\prime} j^\mu \phi^\dagger_v \mid 0 > .
\eqno(3.40)
$$

\noindent Setting

$$
M_{\mu \nu} = i \delta \epsilon_{\mu \nu \alpha \beta} k^\alpha v^\beta ,
\eqno(3.41)
$$

\noindent we obtain

$$
<B_6 (v^\prime , s^\prime ) \mid j_\mu \mid B_{\overline{3}} (v,s) > =
-\frac{\delta}{2 \sqrt{3}} \overline{u} (v^\prime , s^\prime) ( \not{k}
\gamma_\mu - \gamma_\mu \not{k} ) u (v,s) . \eqno(3.42)
$$

\noindent It then follows that

$$
<B_6 (v^\prime , s^\prime) \gamma (k, \epsilon) \mid j_\mu A^\mu \mid
B_{\overline{3}} (v,s) > = \frac{\delta}{2 \sqrt{3}} \overline{u} (v^\prime
, s^\prime) \sigma \cdot F u(v, s) , \eqno(3.43)
$$

\noindent and

$$
a_2 = \frac{1}{2 \sqrt{3}} \delta . \eqno(3.44)
$$

\noindent Likewise, for the $a_4$ coupling we have

$$
<B^\ast_6 (v^\prime , s^\prime) \gamma (k, \epsilon) \mid j_\mu A^\mu \mid
B_{\overline{3}} (v,s) > = \frac{\delta}{2} \epsilon_{\mu \nu \alpha \beta}
\overline{u}^\mu (v^\prime , s^\prime) \gamma^\nu F^{\alpha \beta} u (v,s)
, \eqno(3.45)
$$

\noindent and

$$
a_4 = \frac{\delta}{2} . \eqno(3.46)
$$

\noindent Eqs. (3.24), (3.32), (3.35), (3.38), (3.44) and (3.46) together
give

$$
a_3 = - \frac{\sqrt{3}}{2} a_1 ~,~ a_5 = - \frac{3}{2} a_1  ~,~ a_4 = \sqrt{3}
a_2 ~,~  a_6 = 0 . \eqno(3.47)
$$

\noindent Consequently, only two of the six couplings $a_1 , \cdots a_6$
are independent.  Furthermore, these two couplings are independent of the
heavy masses.

There are two corrections which we would like to
incorporate in the Lagrangian (3.9).  First, when the heavy quark mass is
not infinite, i.e. $m_Q \neq \infty$, we may take into account the effects
of the couplings $a^\prime_i$ induced by heavy quarks and of the Dirac
magnetic moments $\mu_B$ of heavy baryons.  Second, as in the meson case,
SU(3) breaking effects due to light quark mass differences can be
incorporated by replacing the charge matrix ${\cal{Q}}$ by

$$
{\cal{Q}} \rightarrow ~\stackrel{\sim}{\cal{Q}} = \left ( \begin{array}{ccc}
\frac{2}{3} & 0 & 0 \\ 0 & -\frac{\alpha}{3} & 0 \\ 0 & 0 &  -
\frac{\beta}{3}
\end{array} \right) , \eqno(3.48)
$$

\noindent where

$$
\alpha = \frac{m_u}{m_d}  ~,~~~ \beta = \frac{m_u}{m_s} . \eqno(3.49)
$$

\noindent This is equivalent to adding Lagrangian terms like

$$
\Delta {\cal{L}} =
\left( \frac{1}{m_d} - \frac{1}{m_u} \right) {\rm tr} \left(
\overline{B}^\prime {\cal{Q}} \sigma \cdot F B  \right)~{\rm or}~~\left(
\frac{1}{m_s} - \frac{1}{m_u} \right) {\rm tr} \left( \overline{B}^\prime
{\cal{Q}} \sigma \cdot F B \right) , \eqno(3.50)
$$

\noindent to (3.9).

We now use the nonrelativistic quark model to calculate the coupling
constants $a_i$ and $a^\prime_i$.  We choose the magnetic field along the
$z$ direction so that

$$
\sigma_{\mu \nu} F^{\mu \nu} = - 2 \stackrel{\rightharpoonup}{\sigma} \cdot
\stackrel{\rightharpoonup}{H} = - 2 \sigma_z H .  \eqno(3.51)
$$

\noindent Note that in the rest frame of the heavy baryon

$$
\begin{array}{rcl}
\epsilon_{\mu \nu \alpha \beta} \overline{B}^{\ast \mu}_6
\stackrel{\sim}{\cal{Q}} \gamma^\nu F^{\alpha \beta} B_6 & = & 2 \left(
\stackrel{\rightharpoonup}{B}^\ast_6 \right)_z \stackrel{\sim}{\cal{Q}} B_6
H ~~, \\ ~~ \\
\overline{B}^{\ast \mu}_6 \stackrel{\sim}{\cal{Q}} F_{\mu \nu} B^{\ast
\nu}_6 & = & - \left( \stackrel{\rightharpoonup}{B}^\ast_6
\stackrel{\sim}{\cal{Q}} \times \stackrel{\rightharpoonup}{B}^\ast_6
\right)_z H~~,
\end{array}
\eqno(3.52)
$$

\noindent and the wave function of the
$\stackrel{\rightharpoonup}{B}^\ast_6$ is given by

$$
\begin{array}{lcl}
\stackrel{\rightharpoonup}{B}^\ast_6 \left( \frac{3}{2} \right) &  =  &
\stackrel{\rightharpoonup}{\varepsilon}_1 u_\uparrow ~~, ~~~~~{\rm for}~ s_z
= \frac{3}{2} ~~, \\ ~~ \\
\stackrel{\rightharpoonup}{B}^\ast_6 \left( \frac{1}{2} \right) & = &
\frac{1}{\sqrt{3}}
\stackrel{\rightharpoonup}{\varepsilon}_1 u_\downarrow + \sqrt{\frac{2}{3}}
\stackrel{\rightharpoonup}{\varepsilon}_3 u_\uparrow ~, ~~ {\rm for}~ s_z =
\frac{1}{2} ~~,
\end{array}
\eqno(3.53)
$$

\noindent where

$$
\stackrel{\rightharpoonup}{\varepsilon_1} = - \frac{1}{\sqrt{2}} (1, i, 0)
 ~,~~~ \stackrel{\rightharpoonup}{\varepsilon_3} = (0, 0, 1) .  \eqno(3.54)
$$

\noindent By working out the trace terms tr$(\overline{B}^\prime
\stackrel{\sim}{\cal{Q}} B )$ for $B= B_{\overline{3}}~,~B_6$ and
$B^\ast_6$, we obtain

$$
\begin{array}{l}
<\Sigma^{+1}_Q \uparrow \mid {\cal{L}}^{(2)}_B \mid \Sigma^{+1}_Q \uparrow
> = -2 ( \frac{2}{3} a_1 + e_Q a^\prime_1 ) ~, \\ ~~ \\
< \Sigma^0_Q \uparrow \mid {\cal{L}}^{(2)}_B \mid \Lambda_Q \uparrow > = -
\sqrt{2} a_2 (\frac{2}{3} + \frac{\alpha}{3} ) ~, \\ ~~ \\
< \Sigma^{\ast +1}_Q (\frac{1}{2}) \mid {\cal{L}}^{(2)}_B \mid \Sigma^{+1}_Q
\uparrow > = 2 \sqrt{\frac{2}{3}} ( \frac{2}{3} a_3 + e_Q a^\prime_3 ) ~,
\\ ~~ \\
<\Sigma^{\ast 0}_Q (\frac{1}{2}) \mid {\cal{L}}^{(2)}_B \mid \Lambda_Q
\uparrow > = 2 \sqrt{\frac{2}{3}} [ \frac{1}{\sqrt{2}} (\frac{2}{3} +
\frac{\alpha}{3} ) a_4 ] ~, \\ ~~ \\
< \Sigma^{\ast +1}_Q (\frac{3}{2}) \mid {\cal{L}}^{(2)}_B \mid \Sigma^{\ast
+1}_Q (\frac{3}{2}) > = 2 ( \frac{2}{3} a_5 + e_Q a^\prime_5 ) ~,
\\~~ \\
< \Lambda_Q \uparrow \mid {\cal{L}}^{(2)}_B \mid \Lambda_Q \uparrow > = - 2
[ (\frac{2}{3} - \frac{\alpha}{3}) a_6 + 2 e_Q a^\prime_6 ] ~,
\end{array}
\eqno(3.55)
$$
\noindent where we have dropped the magnetic field $H$ for convenience.
The number in parentheses after a $B^\ast_6$ state indicates the value of
$s_z$.

In the quark model the spin-flip magnetic interaction has the form
$$
{\cal{L}}_{\rm em} = \stackrel{\rightharpoonup}{\mu} \cdot
\stackrel{\rightharpoonup}{H} , \eqno(3.56)
$$
with
$$
\stackrel{\rightharpoonup}{\mu} = \sum_{q} ~~ \mu_q
\stackrel{\rightharpoonup}{\sigma}_q ,
\eqno(3.57)
$$
where $\mu_q = e_q \frac{e}{2m_q}$ is the magnetic moment of the
quark $q$ with its electric charge $e_q$ in units of $e$.  Next, the
flavor-spin wave functions of heavy baryons needed are
$$
\begin{array}{l}
\mid \Sigma^{+1}_Q \uparrow > = \frac{1}{\sqrt{6}} \bigg[2 \mid Q \downarrow >
\mid u \uparrow u \uparrow > - \mid Q \uparrow > ( \mid u \uparrow u
\downarrow > + \mid u \downarrow u \uparrow > ) \bigg] ~, \\ ~~ \\
\mid \Sigma^0_Q \uparrow > = \frac{1}{\sqrt{12}} \bigg[ 2 \mid Q \downarrow >
(\mid u \uparrow d \uparrow > + \mid d \uparrow u \uparrow > ) \\ ~~ \\
 ~~~~~~~~~- \mid Q
\uparrow > ( \mid u \uparrow d \downarrow > + \mid d \uparrow u \downarrow
> + \mid u \downarrow d \uparrow > +  \mid d \downarrow u \uparrow > ) \bigg]~,
\\ ~~ \\
\mid \Lambda_Q \uparrow> = \frac{1}{2} \mid Q \uparrow > ( \mid u \uparrow d
\downarrow > - \mid u \downarrow d \uparrow > - \mid d \uparrow u \downarrow
> + \mid d \downarrow u \uparrow > )~, \\ ~~ \\
\mid \Sigma^{+1 \ast} ( \frac{1}{2}) > = \frac{1}{\sqrt{3}} \bigg[ \mid Q
\downarrow > \mid u \uparrow u \uparrow > + \mid Q \uparrow > ( \mid u
\uparrow u \downarrow > + \mid u \downarrow u \uparrow > ) \bigg]~, \\ ~~ \\
\mid \Sigma^{+1 \ast}_Q (\frac{3}{2}) > = \mid Q \uparrow > \mid u \uparrow
u \uparrow > ~, \\ ~~ \\
\mid \Sigma^{0 \ast}_Q ( \frac{1}{2} ) > = \frac{1}{\sqrt{6}} \bigg[ \mid Q
\downarrow > ( \mid u \uparrow d \uparrow > + \mid d \uparrow u \uparrow >
) \\ ~~ \\  ~~~~~~~~~
+ \mid Q \uparrow > (\mid u \uparrow d \downarrow > + \mid u \downarrow d
\uparrow > + \mid d \uparrow u \downarrow > + \mid d \downarrow u \uparrow
> ) \bigg]~.
  \end{array}   \eqno(3.58)
$$

It is then straightforward to show that

$$
\begin{array}{l}
<\Sigma^{+1}_Q \uparrow \mid {\cal{L}}_{\rm em} \mid \Sigma^{+1}_Q \uparrow
> = \frac{4}{3} \mu_u - \frac{1}{3} \mu_Q ~, \\ ~~ \\
< \Sigma^0_Q \uparrow \mid {\cal{L}}_{\rm em} \mid \Lambda_Q \uparrow > = -
\frac{1}{2 \sqrt{3}} \mu_u (2+ \alpha )~, \\ ~~ \\
< \Lambda_Q \uparrow \mid {\cal{L}}_{\rm em} \mid \Lambda_Q \uparrow > =
\mu_Q~, \\ ~~ \\
< \Sigma^{+1 \ast}_Q ( \frac{1}{2}) \mid {\cal{L}}_{\rm em} \mid
\Sigma^{+1}_Q \uparrow > = \frac{2 \sqrt{2}}{3} ( \mu_u - \mu_Q ) ~, \\ ~~
\\
< \Sigma^{0 \ast}_Q ( \frac{1}{2}) \mid {\cal{L}}_{\rm em} \mid \Lambda_Q
\uparrow > = \frac{1}{\sqrt{6}} \mu_u (2 + \alpha)~, \\ ~~ \\
< \Sigma^{+1 \ast}_Q (\frac{3}{2}) \mid {\cal{L}}_{\rm em} \mid \Sigma^{+1
\ast}_Q (\frac{3}{2}) > = 2 \mu_u + \mu_Q~,
   \end{array}   \eqno(3.59)
$$

\noindent where we have dropped the magnetic field as before.  Comparing
this with Eq. (3.55) leads to

$$
\begin{array}{c}
a_1 = - \mu_u ~~,~~ a_2 = \frac{\sqrt{6}}{4} \mu_u ~~,~~ a_3 =
\frac{\sqrt{3}}{2} \mu_u ~~,~~ \\ ~~ \\
a_4 = \frac{3}{2 \sqrt{2}} \mu_u ~~,~~ a_5 = \frac{3}{2} \mu_u ~~,~~
a_6 = 0 ,  \end{array}
 \eqno(3.60)
$$

\noindent and

$$
\begin{array}{c}
a^\prime_1 = \frac{1}{6} \frac{\mu_Q}{e_Q} ~~,~~ a^\prime_2 = a^\prime_4 =
0 ~~,~~ a^\prime_3 = -\frac{1}{\sqrt{3}} \frac{\mu_Q}{e_Q} , \\ ~~ \\
a^\prime_5 = \frac{1}{2} \frac{\mu_Q}{e_Q} ~~,~~ a^\prime_6 = - \frac{1}{4}
\frac{\mu_Q}{e_Q}  . \end{array}
\eqno(3.61)
$$

\noindent It is evident that the relations (3.47) for the couplings $a_1
\cdots a_6$ predicted by the heavy quark symmetry are satisfied in the quark
model calculation, as they should be.

\bigskip\bigskip
\noindent{\bf IV.~~Applications}
\bigskip

In this section we apply our results obtained so far to the electromagnetic
decays of the heavy hadrons.  As we recall, there are six unknown coupling
constants in the baryon sector, but they are reduced to two via the use of
heavy quark symmetry.  The nonrelativistic quark model is then applied to
compute them.  Consequently, the dynamics of the radiative transitions for
emission of soft photons and pions is completely determined by the heavy
quark symmetry and chiral symmetry, supplemented by the quark model.

As an application, we first focus on the two-body radiative decays such as
$P^\ast \rightarrow P \gamma ,~\Sigma_Q \rightarrow \Lambda_Q \gamma ,~
\Xi^\prime_Q \rightarrow \Xi_Q \gamma$.  Since the heavy hadrons, e.g.,
$B^\ast,~\Xi^\prime_c ,~\Xi_b$ are dominated by the electromagnetic decays,
the decay widths of these heavy particles can be directly calculated.  When
combining with our previous results [3] on the strong decays of $D^\ast$ and
$\Sigma_c$, we can also predict the total widths and branching ratios of
these particles.  We shall also consider the radiative decays involving one
pion emission.  Some examples of kinematically allowed modes are $\Sigma_c
\rightarrow \Lambda_c \pi \gamma, \Sigma^\ast_c \rightarrow \Lambda_c \pi
\gamma , \Xi^\ast_c \rightarrow \Xi_c \pi \gamma$, etc.  We shall see later
that the decay $\Sigma^+_c \rightarrow \Lambda^+_c \pi^0 \gamma$ provides a
nice test on the chiral structure of the electromagnetic gauge invariant
Lagrangian ${\cal{L}}^{(2)}_B$, whereas the four-particle contact
interaction dictated by ${\cal{L}}^{(1)}_B$ can be tested by the other
channel $\Sigma^0_c \rightarrow \Lambda^+_c \pi^- \gamma$.

We begin with the $P^\ast \rightarrow P \gamma$ decays.  The decay width
corresponding to the general amplitude

$$
A \left[ P^\ast (v, \epsilon^\ast ) \rightarrow P \gamma (k, \epsilon)
\right] = -i \rho \epsilon_{\mu \nu \alpha \beta} k^\mu \epsilon^\nu
v^\alpha \epsilon^{\ast \beta} ,
\eqno(4.1)
$$

\noindent is

$$
\Gamma (P^\ast \rightarrow P \gamma ) = \frac{\rho^2}{12 \pi M^{\ast 2}}
k^3 , \eqno(4.2)
$$

\noindent where $k$ is the photon momentum in the CM system.  From Eqs.
(2.19), (2.41) and (2.42) we obtain the couplings

$$
\begin{array}{cccc}
\rho (P^{\ast \frac{1}{2}}) & = & 2 \sqrt{M_P M_{P^\ast}} &
\left(-\frac{1}{3}
\frac{e}{2m_d} + e_Q \frac{e}{2m_Q} \right) , \\ ~~ \\
\rho (P^{\ast -\frac{1}{2}}) & = & 2 \sqrt{M_P M_{P^\ast}} & \left(
\frac{2}{3} \frac{e}{2m_u} + e_Q \frac{e}{2m_Q} \right) , \\ ~~ \\
\rho (P^{\ast 0}) & = & 2 \sqrt{M_P M_{P^\ast}} & \left( -\frac{1}{3}
\frac{e}{2m_s} + e_Q \frac{e}{2m_Q} \right) .
\end{array}
\eqno(4.3)
$$

\noindent As an example, the computed results for $D^\ast \rightarrow D +
\gamma $ are exhibited in Table I for the constituent quark masses

$$
m_u = 338\, {\rm MeV},~~~~m_d = 322\, {\rm MeV},~~~~m_s = 510\, {\rm MeV},
\eqno(4.4)
$$

\noindent given by the Particle Data Group [10], and $m_c = 1.6$ GeV.  To
determine the $D^\ast$ branching ratios we have included the partial widths
of $D^\ast \rightarrow D \pi$ predicted in Ref. [3] with the axial quark
coupling $g^{ud}_A = 0.75$.  It is evident that the agreement between theory
and the most recent experimental measurement of CLEO II [9] is excellent.
In
particular, the observed small branching ratio of $D^{\ast +} \rightarrow
D^+ \gamma$ by CLEO II is consistent with our theoretical expectation,
contrary
to the large PDG [10] average value.  This also means that it is not
necessary to invoke a large anomalous magnetic moment for the charm quark
as previously conjectured.  The total widths of the $D^\ast$ [12] are

$$
\begin{array}{c}
\Gamma_{\rm tot} (D^{\ast +}) = 141\, {\rm keV}, \\ ~~ \\
\Gamma_{\rm tot} (D^{\ast 0}) = 102\, {\rm keV}, \\ ~~ \\
\Gamma_{\rm tot} (D^{\ast +}_s ) = 0.3\, {\rm keV} ,
  \end{array}    \eqno(4.5)
$$

\noindent which are also listed in Table I.  The $\Gamma_{\rm tot} (D^{\ast
+})$ predicted here is very close to the upper limit $\Gamma_{\rm tot}
(D^{\ast +}) < 131 $ keV (90\% CL) published by the ACCMOR Collaboration
[11].  We urge the experimentalists to perform more precision measurements
of $\Gamma_{\rm tot} (D^\ast )$.

\def\ri{\rightarrow}
\begin{table*}[t]
\begin{center}
\begin{tabular}{lccccccc}
\multicolumn{8}{l}{Table~I. The predicted branching ratios of the $D^*$
mesons. The}\\
\multicolumn{8}{l}{\hbox to 48 pt{\hfil}predicted partial widths of $D^*\ri
 D+\pi$ are taken from} \\
\multicolumn{8}{l}{\hbox to 48 pt{\hfil}Ref.[3] for $g_A^{ud}=0.75$. For
comparison, the experimental}\\
\multicolumn{8}{l}{\hbox to 48 pt{\hfil}results of CLEO II [9] and PDG
(1992) [10] average}\\
\multicolumn{8}{l}{\hbox to 48 pt{\hfil}values are given in the last two
columns.}\\
\hline\hline
decay mode &$\Gamma$(keV) & Br(\%)$_{\rm theory}$ & Br($\%)_{\rm CLEO}$
& Br$(\%)_{\rm PDG}$ \\
\noalign{\vspace{2pt}}
\hline
\noalign{\vspace{2pt}}
$D^{*+}\ri D^0\pi^+$ & 95 & 67.3 & $68.1\pm 1.0\pm 1.3$ & $55\pm 4$ \\
$D^{*+}\ri D^+\pi^0$ & 44 & 31.2 & $30.8\pm 0.4\pm 0.8$ & $27.2\pm 2.5$ \\
$D^{*+}\ri D^+\gamma$ & 2 & 1.5 & $1.1\pm 1.4\pm 1.6$ & $18\pm 4$ \\
$D^{*+}\ri {\rm all}$ & 141 & & &  \\
\hline
$D^{*0}\ri D^0\pi^0$ & 68 & 66.7 & $63.6\pm 2.3\pm 3.3$ & $55\pm 6$ \\
$D^{*0}\ri D^0\gamma$ & 34 & 33.3 & $36.4\pm 2.3\pm 3.3$ & $45\pm 6$ \\
$D^{*0}\ri {\rm all}$ & 102 & & & \\
\hline
$D^{*+}_s\ri D^+_s\gamma$ & 0.3 & $\sim$100 & & \\
 \hline\hline
\end{tabular}
\end{center}
\end{table*}

Before proceeding, we should stress that it is important to include the
corrections due to the magnetic moment of the charm quark as its mass is
not too large compared to the light quarks, $m_s / m_c \approx \frac{1}{3}$
and its charge is $\frac{2}{3} e$.  It is clear from Eq. (4.3) that the
charm-quark contribution is largely destructive in the radiative decays of
$D^{\ast +}$ and $D^{\ast +}_s$.  Had we worked in the heavy quark limit,
we would have obtained

$$
\begin{array}{c}
\Gamma (D^{\ast 0} \rightarrow D^0 \gamma ) = 23\, {\rm keV}, \\ ~~ \\
\Gamma (D^{\ast +} \rightarrow D^+ \gamma ) = 6\, {\rm keV}, \\ ~~ \\
\Gamma (D^{\ast +}_s \rightarrow D^+_s \gamma ) = 2.4\, {\rm keV},
  \end{array} \eqno(4.6)
$$

\noindent which are significantly different from those presented in Table
I.  Finally, for completeness we also give the results for the radiative
decays of $B^\ast$

$$
\begin{array}{c} \Gamma (B^{\ast +}_u \rightarrow B^+_u \gamma )= 0.84\,{\rm
 keV},
\\ ~~ \\ \Gamma (B^{\ast 0}_d \rightarrow B^0_d \gamma ) = 0.28\,{\rm keV},
\end{array}
\eqno(4.7)
$$

\noindent where we have used the mass values $m_{B^\ast} = 5324.6$ MeV and
$m_b = 5$ GeV.

We next turn to the baryon sector and consider the following two-body
radiative decays with some specific examples

$$
\begin{array}{ccccc}
B_6 \rightarrow B_{\overline{3}} + \gamma & : & \Sigma_Q \rightarrow
\Lambda_Q + \gamma, & \Xi^\prime_Q \rightarrow \Xi_Q + \gamma , & \\ ~~ \\
 B^\ast_6 \rightarrow B_{\overline{3}} + \gamma & : & \Sigma^\ast_Q
\rightarrow \Lambda_Q + \gamma , & \Xi^{\prime \ast}_Q \rightarrow \Xi_Q +
\gamma , & \\ ~~ \\
B^\ast_6 \rightarrow B_6 + \gamma & : & \Sigma^\ast_Q \rightarrow \Sigma_Q
+ \gamma , & \Xi^{\prime \ast}_Q \rightarrow \Xi^\prime_Q + \gamma , &
\Omega^\ast_Q \rightarrow \Omega_Q + \gamma .
\end{array}
\eqno(4.8)
$$
The electromagnetic decay of a sextet baryon $B_6$ into a
$B_{\overline{3}}$ plus a photon is described by the amplitude

$$
M \left( B_6 \rightarrow B_{\overline{3}} + \gamma (k) \right) = i \eta_1
\overline{u}_{\overline{3}}  \sigma_{\mu \nu} k^\mu \epsilon^\nu u_6 ~.
\eqno(4.9)
$$

\noindent Its decay rate is simply given by

$$
\Gamma (B_6 \rightarrow B_{\overline{3}} + \gamma ) = \frac{1}{\pi}
\eta^2_1 k^3 ,
\eqno(4.10)
$$

\noindent where $k$ is the photon momentum in the CM system.  For
completeness, we give here the results of the radiative decay of a spin
$\frac{3}{2}$ heavy baryon, though none of these heavy baryons have been
found yet.  The amplitude of the transition $B^\ast_6 \rightarrow
B_{\overline{3}} + \gamma$ reads

$$
M (B^\ast_6 \rightarrow B_{\overline{3}} + \gamma ) = i \eta_{2}
\epsilon_{\mu \nu \alpha \beta} \overline{u} \gamma^\nu k^\alpha
\epsilon^\beta u^\mu . \eqno(4.11)
$$

\noindent The evaluation of the corresponding decay width involves the use
of the projection operator

$$
\begin{array}{lc}
P_{\mu \nu} (v) \equiv &
\sum_s u_\mu (p,s) \overline{u}_\nu (p,s) = \frac{(\not{p} + m)}{2m} \left[
-g_{\mu \nu} + \frac{1}{3} \gamma_\mu \gamma_\nu + \frac{1}{3m} (\gamma_\mu
p_\nu - \gamma_\nu p_\mu) + \frac{2}{3m^2} p_\mu p_\nu \right] \\ ~~ \\
 & = \left[ -g_{\mu \nu} + \frac{1}{3} \gamma_\mu \gamma_\nu - \frac{1}{3m}
(\gamma_\mu p_\nu - \gamma_\nu p_\mu ) + \frac{2}{3m^2} p_\mu p_\nu \right]
\frac{(\not{p} + m)}{2m} . \end{array}  \eqno(4.12)
$$

\noindent The final result is

$$
\Gamma (B^\ast_6 \rightarrow B_{\overline{3}} + \gamma ) = \frac{k}{48 \pi}
\eta^2_2 \left( 1 - \frac{m^2_f}{m^2_i} \right)^2 (3 m^2_i + m^2_f ) ,
\eqno(4.13)
$$

\noindent where $m_i~~ (m_f)$ is the mass of the initial (final) baryon in
the decay.  Except for a different coupling constant, a similar formula holds
for the decay $B^\ast_6 \rightarrow B_6 + \gamma$.

We are ready to elaborate on the above results by some examples.  The first
example is $\Sigma^+_c \rightarrow \Lambda^+_c + \gamma$.  From Eqs. (3.9),
(3.48), (3.49) and (3.60) we find

$$
\begin{array}{ccc}
\eta_1 ( \Sigma^+_c - \Lambda^+_c ) & = & \sqrt{2} \left( \frac{2}{3}
+ \frac{\alpha}{3} \right) a_2 \\ ~~ \\
 & = & \frac{1}{\sqrt{3}} \frac{e}{2m_u} \left( \frac{2}{3} + \frac{1}{3}
\frac{m_u}{m_d} \right) , \end{array}
\eqno(4.14)
$$

\noindent which in turn implies that

$$
\Gamma \left( \Sigma^+_c \rightarrow \Lambda^+_c + \gamma \right) =
93\,{\rm keV} .
\eqno(4.15)
$$

\noindent This together with the partial rate $\Gamma ( \Sigma^+_c
\rightarrow \Lambda^+_c \pi^0 ) = 2.43\,{\rm MeV}$  (for $g^{ud}_A = 0.75$)
obtained in Ref. [3] yields the total decay width of $\Sigma^+_c$

$$
\Gamma_{\rm tot} ( \Sigma^+_c ) = 2.54\, {\rm MeV} , \eqno(4.16)
$$

\noindent and the branching ratio of $\Sigma^+_c \rightarrow \Lambda^+_c +
\gamma$

$$
{\rm Br} ( \Sigma^+_c \rightarrow \Lambda^+_c + \gamma ) = 3.8 \% .
\eqno(4.17)
$$

\noindent The second example is $\Xi^\prime_c \rightarrow \Xi_c + \gamma$.
The coupling $\eta_1$ is given by

$$
\begin{array}{ccccc}
\eta_1 (\Xi^{\prime +}_c - \Xi^+_c ) & = & \sqrt{2} \left( \frac{2}{3} +
\frac{\beta}{3} \right) a_2 & = & \frac{1}{\sqrt{3}} \frac{e}{2m_u} \left(
\frac{2}{3} + \frac{m_u}{3m_s} \right) , \\ ~~ \\
\eta_1 (\Xi^{\prime 0 }_c - \Xi^0_c ) & = & \sqrt{2} \left( -
\frac{\alpha}{3} + \frac{\beta}{3} \right) a_2 & = & \frac{e}{6 \sqrt{3}}
\left( \frac{1}{m_s} - \frac{1}{m_d} \right) , \end{array}
\eqno(4.18)
$$

\noindent for $\Xi^{\prime +}_c - \Xi^+_c$ and $\Xi^{\prime 0}_c - \Xi^0_c
$ transitions, respectively.  We get

$$
\Gamma (\Xi^{\prime +}_c \rightarrow \Xi^+_c + \gamma ) = 16\, {\rm
keV}~,~~\Gamma (\Xi^{\prime 0}_c \rightarrow \Xi^0_c + \gamma ) = 0.3\, {\rm
keV} . \eqno(4.19)
$$

\noindent In the above we have used the mass $m_{\Xi_c} = 2470$ MeV from
PDG (1992) [10], and the mass difference $m_{\Xi^\prime_c} - m_{\Xi_c}
\simeq 100 $ MeV from a theoretical estimate [13].  We also assume no
mixing between $\Xi^\prime_c$ and $\Xi_c$.  If the mass difference turns
out to be this small, there will be no strong decays for $\Xi^\prime_c$.
We thus have a prediction for the total width of $\Xi^\prime_c$:

$$
\Gamma_{\rm tot} ( \Xi^{\prime +}_c ) = 16\, {\rm keV}~,~~\Gamma_{\rm tot}
(\Xi^{\prime 0}_c ) = 0.3\, {\rm keV} .  \eqno(4.20)
$$

\noindent So far, the examples of radiative decays considered do not test
critically the heavy quark symmetry nor the chiral symmetry.  The results
follow simply from the quark model.  We now offer examples in which both
the heavy quark symmetry and the chiral symmetry enter in a crucial way.
These are the radiative decays of heavy baryons involving an emitted pion.
Some examples which are kinematically allowed are

$$
\Sigma_c \rightarrow \Lambda_c \pi \gamma ~~,~~ \Sigma^\ast_c \rightarrow
\Lambda_c \pi \gamma ~~,~~ \Sigma^\ast_c \rightarrow \Sigma_c \pi
\gamma~~,~~  \Xi^\ast_c \rightarrow \Xi_c \pi \gamma
$$

\noindent To be specific, we focus on the decay $\Sigma_c \rightarrow
\Lambda_c \pi \gamma$.  The Feynman diagrams for the decay follow from the
Lagrangian ${\cal{L}}^{(1)}_B$ and ${\cal{L}}^{(2)}_B$ given in the last
section.  There are a total of eight possible diagrams as depicted in Fig.
1:
six of them arise from baryon poles, one from the meson pole, and one from
the four-point contact term.  For the present discussion, we will limit
ourselves to the situation in the heavy quark limit as to bring out the
simplifications that occur in the symmetry limit.  Thus, the pion and the
photon are both soft, and we will neglect terms of order $\frac{q}{m_Q}$
and/or $\frac{k}{m_Q}$ with $q$ and $k$ being the pion and photon momentum,
respectively.  It turns out that the contact interaction dictated by the
Lagrangian ${\cal{L}}^{(1)}_B$ can be nicely tested by the decay $\Sigma^0_c
\rightarrow \Lambda^+_c \pi^- \gamma$, whereas a test on the chiral
structure of ${\cal{L}}^{(2)}_B$ is provided by the process $\Sigma^+_c
\rightarrow \Lambda^+_c \pi^0 \gamma$.  Let us discuss the latter first.

It is interesting to see that only diagrams (d) and (f) survive in the
heavy quark limit.  Diagrams (b) and (c) vanish because of isospin
conservation.  Diagrams (g) and (h) do not exist for a neutral pion.
Diagram (e) is prohibited owing to the absence of the $B_{\overline{3}}
B_{\overline{3}} \pi$ coupling.  The $\Lambda_c \Lambda_c \gamma$ coupling
of diagram (a) is of the convection current type only [cf. Eq. (3.24)] and
in the heavy quark limit it is cancelled out by a similar convection
current $\Sigma_c \Sigma_c \gamma$ coupling of diagram (d).  [This
cancellation is also required by gauge invariance.]  Consequently, we only
have to consider diagram (f) and the magnetic coupling of diagram (d).  The
amplitudes are

$$
\begin{array}{l}
A(\Sigma^+_c \rightarrow \Lambda^+_c \pi^0 \gamma ) = A_d + A_f \\ ~~ \\
A_d = i \frac{4 \sqrt{2} a_1 g_2}{3 f_\pi} \frac{1}{v \cdot k}
\overline{u}_{\Lambda_c} (v^\prime, s^\prime ) (\not{q} - q \cdot v )
\sigma_{\mu \nu}
k^\mu \epsilon^\nu u_{\Sigma_c} (v, s) \\ ~~ \\
A_f = - i \frac{a_3 g_4}{\sqrt{2} f \pi} \left( \frac{2}{3} -
\frac{\alpha}{3} \right) \frac{1}{-v \cdot k + M_{\Sigma_c} -
M_{\Sigma^\ast_c}} \overline{u}_{\Lambda_c} (v^\prime , s^\prime) q_\sigma
P^{\sigma \lambda} (v^\prime) \epsilon_{\lambda \nu \alpha \beta}
\gamma^\nu k^\alpha \epsilon^\beta u_{\Sigma_c} (v, s) \end{array} \eqno(4.21)
$$

\noindent where $P^{\sigma \lambda} (v^\prime)$ is the projection operator
given by (4.12).  Recall that

$$
g_4 = -\sqrt{3} g_2~,~~ g_2 = - 0.75 \sqrt{\frac{2}{3}}  \eqno(4.22)
$$

\noindent for $g^{ud}_A = 0.75$.  Beyond the heavy quark limit, obvious
$1/m_Q$ corrections arise from the magnetic moment $\mu_c$ of the charm
quark and $\mu_B$ of the charmed baryons.

We now come back to the decay $\Sigma^0_c \rightarrow \Lambda^+_c \pi^-
\gamma$.  The main contribution comes from the convection current coupling
of diagrams (a), (g), and (h).  Other diagrams due to the magnetic-type
couplings are suppressed by factors of $k/m_u$, which should be small since
$m_{\Sigma_c} - m_{\Lambda_c} - m_\pi \sim 30$ MeV, and $m_u \sim 330$ MeV.
The contact-term Lagrangian for diagram (g) can be read off from Eqs.
(3.8) and (2.11b),

$$
{\cal{L}}^\prime = -\frac{ieg_2}{\sqrt{2} f_\pi} A_\mu
\overline{\Lambda}^+_c \gamma^\mu \gamma_5 \pi^+ \Sigma_c^0 . \eqno(4.23)
$$

\noindent The amplitudes are given by

$$
A (\Sigma^0_c \rightarrow \Lambda^+_c \pi^- \gamma ) \cong A_a + A_g + A_h
, \eqno(4.24)
$$

\noindent with

$$
\begin{array}{lcl}
A_a  & = & \frac{eg_2}{\sqrt{2} f_\pi} \overline{u}_{\Lambda_c} v \cdot
\epsilon \frac{1}{v \cdot k} \not{q} \gamma_5 u_{\Sigma_c} , \\ ~~ \\
A_g & = & \frac{eg_2}{\sqrt{2} f_\pi} \overline{u}_{\Lambda_c} \not\epsilon
\gamma_5 u_{\Sigma_c}, \\ ~~ \\
A_h & = & -\frac{eg_2}{\sqrt{2} f_\pi} \frac{q \cdot \epsilon}{q \cdot k}
\overline{u}_{\Lambda_c} (\not{q} + \not{k} ) \gamma_5 u_{\Sigma_c} .
\end{array} \eqno(4.25)
$$

\noindent It is easily seen that gauge invariance is respected.  It will be
interesting to work out the energy and angular distributions of the pion
and the $\Lambda_c$.  A detailed analysis of this will be presented in a
future publication.

Aside from the decay rates for $B^\ast \rightarrow B \gamma$ given by
(4.7), we have not calculated any of the radiative decay rates for baryons
containing a $b$-quark.  This is only because there is scarcely any data on
the masses and mass differences of these baryons.  Once they are known, the
same equations (3.59)~--~(3.61) and (4.9)~--~(4.13) can be applied to obtain
the decay rates.

\bigskip\bigskip
\noindent{\bf V.~~Conclusions}
\bigskip

The heavy quark symmetry and the chiral symmetry together provide an ideal
framework for studying the low energy dynamics of heavy mesons and heavy
baryons.  Symmetry considerations reduce to a minimum the number of free
parameters in the theory, and symmetry breaking corrections can be
estimated in principle.  Yet, few if any quantitative predictions can be
made in strong and electromagnetic interactions without further
assumptions.  It is here that the nonrelativistic quark model comes to the
rescue.  All the free parameters needed for the low energy dynamics of
ground state heavy hadrons are calculable in the nonrelativistic quark
model.  Moreover, these calculations depend only on the spin-flavor wave
functions of the quarks, and are independent of the details of the spatial
wave functions.  Therefore, simplicity and (almost) uniqueness characterize
these quark model predictions.  We regard them as a theoretical benchmark
to be compared with experiments as well as other theoretical models.

In Refs.[3], [7] and present work we have explored in detail the
predictions of this theoretical formalism on strong decays, heavy flavor
conserving nonleptonic decays, and radiative decays of the heavy hadrons.
These results may now be combined to obtain predictions for the total
widths and branching ratios of certain heavy particles.  In particular, the
branching ratios obtained for $D^{\ast +}$ and $D^{\ast 0}$ agree very well
with the most recent measurements of CLEO II.  This excellent agreement
between theory and experiment makes it ever more urgent to study and
understand the various symmetry breaking corrections to the strong and
radiative decays.  This is particularly so should the upper limit for
$\Gamma_{\rm tot} (D^{\ast +})$ [11] be confirmed by future experiments.
We would like to know if it is possible to incorporate these corrections to
improve the quark model calculations.  We have begun an investigation to
answer these questions.  The $1/m_Q$ corrections due to the heavy quark's
magnetic moment that we have included in Sections II and III are exact as a
result of the normalization conditions of the Isgur-Wise functions at $v =
v^\prime$.  Other $1/m_Q$ corrections including those to the light quarks'
electromagnetic currents and axial vector currents require a more careful
discussion.  We will communicate these results in a future publication.

There are many other weak radiative decay modes of great interest like

$$
B \rightarrow D (D^\ast) \gamma~,~~~\Lambda_b \rightarrow \Sigma_c
\gamma~,~~~\Xi_b \rightarrow \Xi_c \gamma~~.
$$

\noindent Unfortunately, the effective heavy quark theory developed thus
far cannot be applied to these processes.  The intermediate states in the
relevant pole diagrams are very far from their mass shell.  For example,
the four-momentum squared of the $D$ pole in the decay $B \rightarrow
D^\ast \gamma$ is $m^2_B$.  This means that the residual momentum of the
$D$ meson defined by $P_\mu = m_D v_\mu + k_\mu$ must be of order $m_B$ so
that the approximation $k/m_D \ll 1$ required by the effective heavy quark
theory is no longer valid.  Nevertheless, there is a special class of weak
radiative decays in which heavy flavor is conserved that deserves a
detailed study.  Some examples are $\Xi_Q \rightarrow \Lambda_Q \gamma$ and
$\Omega_Q \rightarrow \Xi_Q \gamma$.  In these decays, weak radiative
transitions arise from the diquark sector of the heavy baryon whereas the
heavy quark behaves as a ``spectator''.  However, the dynamics of these
radiative decays is more complicated than their counterpart in nonleptonic
weak decays, e.g., $\Xi_Q \rightarrow \Lambda_Q \pi$, which have been
studied in Ref.[7].  We hope to study in the future these heavy flavor
conserving weak radiative decays.

\pagebreak

\centerline{\bf Acknowledgments}
\bigskip

One of us (T.M.Y.) would like to express his deep appreciation of the
hospitality extended to him by the Theory Group at the Institute of
Physics, Academia Sinica, Taipei, Taiwan, ROC during his stay there where
part of the work was done.  We would like to thank Professors D. Cassel, P.
Drell and J. Rosner for useful discussions on Ref.[11].  We also thank Drs.
C.G. Boyd and J.G. Korner for informing us of Ref.[14] and Ref.[15],
respectively, in which the radiative decays of $D^\ast$ are studied by
chiral and heavy quark symmetry.  T.M.Y.'s work is supported in part by the
National Science Foundation.  This research is supported in part by the
National Science Council of ROC under Contract Nos.  NSC81-0208-M-001-04,
NSC81-0208-M-001-06 and NSC81-0208-M-001-54.

\bigskip\bigskip
\bigskip\bigskip
\centerline{\bf Figure Captions}
\bigskip

\noindent{\it Fig.}1.  Possible Feynman diagrams for the decays $\Sigma^+_c
\rightarrow \Lambda^+_c \pi^0 \gamma$ and $\Sigma^0_c \rightarrow
\Lambda^+_c \pi^- \gamma$.

\vfill\eject

\centerline{\bf REFERENCES}

\bigskip\bigskip

\begin{enumerate}

\item N. Isgur and M.B. Wise, {\it Phys. Lett.} {\bf B232}, 113 (1989),
{\it Phys. Lett.} {\bf B237}, 527 (1990).

\item M.B. Voloshin and M.A. Shifman, Yad. Fiz. {\bf 45}, 463 (1987)
[Sov. J. Nucl. Phys. {\bf 45}, 292 (1987)].

\item T.M. Yan, H.Y. Cheng, C.Y. Cheung, G.L. Lin, Y.C. Lin, and H.L.
Yu, {\it Phys. Rev.} {\bf D46}, 1148 (1992).  See also T.M. Yan, to appear
in Chinese J. Phys. ROC.

\item M.B. Wise, {\it Phys. Rev.} {\bf D45}, R2188 (1992).

\item G. Burdman and J. Donoghue, {\it Phys. Lett.} {\bf B280}, 287
(1992).

\item P. Cho, {\it Phys. Lett.} {\bf B285}, 145 (1992).

\item H.Y. Cheng, C.Y. Cheung, G.L. Lin, Y.C. Lin, T.M. Yan and H.L. Yu,
Cornell preprint CLNS 92/1153 and IP-ASTP-07-92, to appear in Phys. Rev.
{\bf D}.

\item Additional applications of the formalism have been made: C. Lee, M.
Lu and M. Wise, CALT-68-1771 (1992); B. Grinstein, E. Jenkins, A. Manohar,
M. Savage and M. Wise, {\it Nucl. Phys.} {\bf B380}, 369 (1992); E. Jenkins
and M. Savage, {\it Phys. Lett.} {\bf B281}, 331 (1992); A. Falk and M.
Luke, SLAC-PUB-5812 (1992); U. Kilian, J. G. Korner and D. Pirjol,
{\it Phys. Lett.} {\bf B288}, 360 (1992); and P. Cho, Harvard preprint
HUTP-92/A039.

\item F. Butler {\it et al}, Cornell preprint 92/1143, CLEO 92-3.

\item Review of Particle Properties, {\it Phys. Rev.} {\bf D45}, S1
(1992).

\item S. Barlarg {\it et al}, (The ACCMOR Collaboration), {\it Phys. Lett.}
{\bf B278}, 480 (1992).

\item Many calculations of the partial and total widths of $D^\ast $
exist in the literature.  We give here only a few references: S. Ono {\it
Phys. Rev. Lett.} {\bf 37}, 655 (1976); E. Eichten {\it et al.}, {\it Phys.
Rev.} {\bf D21}, 203 (1980); W. Wilcox, O.V. Maxwell, and K.A. Milton, {\it
Phys. Rev.} {\bf D31}, 1081 (1985); R.L. Thews and A.N. Kamal, {\it Phys.
Rev.} {\bf D32}, 810 (1985); J. Rosner, {\it Comm. on Nucl. and Part.
Physics}, {\bf 16}, 109 (1986); G.A. Miller and P. Singer, {\it Phys. Rev.}
{\bf D37}, 2564 (1988); A.N. Kamal and Q.P. Xu, {\it Phys. Lett.} {\bf
B284}, 42 (1992); and L. Angelos and G.P. Lepage, {\it Phys. Rev.} {\bf
D45}, 302 (1992).

\item M.J. Savage and M. Wise, {\it Nucl. Phys.} {\bf B326}, 15 (1989).

\item J.F. Amundson, C.G. Boyd, E. Jenkins, M. Luke, A.V. Manohar, J.L.
Rosner, M.J. Savage and M.B. Wise, UCSD/PTH 92-31, CALT-68-1816, EFI-92-45,
CERN-TH 6650/92.  Through this paper, we also became aware of a work on a
similar subject by P. Cho and H. Georgi, Harvard Preprint HUTP-02/A043.

\item J.G. Korner, {\it et al}, Mainz preprint.

\end{enumerate}

\end{document}